\documentclass[superscriptaddress,showpacs,floatfix,preprintnumbers,amsmath,amssymb]{revtex4}

\usepackage{graphicx} 
\usepackage{dcolumn}  
\usepackage{subfigure}
\usepackage{color}
\usepackage{amsmath}
\usepackage{amssymb}

\newcommand{\beq}{\begin{equation}}
\newcommand{\eeq}{\end{equation}}
\newcommand{\beqn}{\begin{eqnarray}}
\newcommand{\eeqn}{\end{eqnarray}}

\newcommand{\qlp}{\ensuremath{q_L^\prime}}
\newcommand{\qrp}{\ensuremath{q_R^\prime}}
\newcommand{\qp}{\ensuremath{q^\prime}}
\newcommand{\blp}{\ensuremath{b_L^\prime}}
\newcommand{\tlp}{\ensuremath{t_L^\prime}}
\newcommand{\brp}{\ensuremath{b_R^\prime}}
\newcommand{\trp}{\ensuremath{t_R^\prime}}
\newcommand{\bp}{\ensuremath{b^\prime}}
\newcommand{\tp}{\ensuremath{t^\prime}}

\newcommand{\Yb}{\ensuremath{Y_{b^\prime}}}
\newcommand{\Yt}{\ensuremath{Y_{t^\prime}}}
\newcommand{\ayb}{\ensuremath{\alpha_{b^\prime}}}
\newcommand{\ayt}{\ensuremath{\alpha_{t^\prime}}}
\newcommand{\at}{\ensuremath{\alpha_W}}
\newcommand{\aon}{\ensuremath{\alpha_Y}}

\newcommand{\half}{\ensuremath{\frac{1}{2}}}
\newcommand{\ont}{\ensuremath{\frac{1}{3}}}
\newcommand{\twt}{\ensuremath{\frac{2}{3}}}
\newcommand{\fot}{\ensuremath{\frac{4}{3}}}
\newcommand{\eit}{\ensuremath{\frac{8}{3}}}

\begin{document}

\affiliation{Institute for Nuclear Research of the Russian Academy of Sciences, 117312 Moscow, Russia}
\affiliation{Institute for Theoretical and Experimental Physics, Moscow 117218, Russia}
\affiliation{Moscow Institute of Physics and Technology, Moscow Region 141700, Russia}
  \author{D.~Gorbunov}\affiliation{Institute for Nuclear Research of the Russian Academy of Sciences, 117312 Moscow, Russia}\affiliation{Moscow Institute of Physics and Technology, Moscow Region 141700, Russia} 
 \author{P.~Pakhlov}\affiliation{Institute for Theoretical and Experimental Physics, Moscow 117218, Russia}\affiliation{Moscow Institute of Physics and Technology, Moscow Region 141700, Russia} 

\title{ \quad\\[0.5cm] \Large Antibaryonic dark matter}

\begin{abstract}
Assuming existence of (very) heavy fourth generation of quarks and
antiquarks we argue that antibaryon composed of the three heavy
antiquarks can be light, stable and invisible, hence a good candidate
for the Dark matter particle.  Such opportunity allows to keep the
baryon number conservation for the generation of the visible baryon
asymmetry. The dark matter particles traveling through the ordinary
matter will annihilate with nucleons inducing
proton(neutron)-decay-like events with $\sim5\,$GeV energy release in
outcoming particles.
\end{abstract}

\pacs{95.35.+d,14.65.Jk,14.20.Pt}

\maketitle
%

\section{Introduction}

One of Sakharov's condition of existence of the Universe as we know
it, is baryon number violation~\cite{Sakh}. It seems indisputable from
the observation of the baryonic matter over antimatter dominance, if
the Universe got nonzero baryon charge at early evolution
stage. However, if the Dark matter of the Universe (DM) is
antibaryonic, and consisting of exactly the same number of antiquarks
as usual matter does, this Sakharov's requirement is no more
obligatory. Such opportunity can be only justified if DM antibaryons
satisfy the following conditions: they are stable (at least at the
Universe lifetime scale) and neutral; their annihilation with normal
matter is an extremely rare phenomenon; moreover, negative results of
direct searches for Dark matter place upper limits on cross section of
DM elastic scattering off baryons.

In this Letter we consider a minor modification of the Standard Model
(SM) where DM can be formed by antibaryons. Namely, we examine the SM
with 4 generations of quarks (SM4)~\footnote{A modification in lepton
  sector needed to cancel the SM gauge anomalies is required but its
  details are irrelevant for our study.}, which was widely discussed
in literature, and argue that baryons built from the heaviest quark
generation (\tp\ and \bp) can be light, stable and invisible. 

With baryon number conserved, there should be a parity of "Light"
baryons and "Dark" antibaryons in the Universe. We thus conclude from
measurement of DM mass density in the Universe (about 5 times higher
than baryonic mass density \cite{PDG}) that "Dark" antibaryons should
have mass of about 5\,GeV. One and the same mechanism --- antibaryon
formation --- is responsible for both DM and baryon asymmetry
generations in the early Universe. It is worth noting that since in
the model the baryon number enters both matter components (baryonic
and dark), it naturally explains the coincidence (within one order of
magnitude) of visible and dark matter contributions to the present
density of the Universe.

The idea of baryon-symmetric Universe \cite{Dodelson:1989ii} was
extensively studied in literature, and several mechanism to generate
antibaryonic dark matter have been suggested, for a recent review see
\cite{Davoudiasl:2012uw}.  In our model a specific production
mechanism does the job, and the phenomenology is different except the
main signature of all antibaryonic dark matter models: an induced
nucleon decay.

\section{Light baryons built of heavy quarks}

Light $u$ and $d$ quarks in the normal baryons (protons and neutrons)
are bound by the strong interaction. For baryons consisting heavy
\tp\ and \bp\ quarks the extra huge binding energy arises due to the
scalar exchange between each pair of quarks. Indeed, scalar fields
provide attractive forces between fermions with the strength
proportional to the product of Yukawa couplings, \emph {i.e.}
increasing with the fermion masses.

The naive (perturbative) estimates of the binding energy of two heavy
quarks $Q$ due to light scalar exchange is
\begin{equation}
E_{\rm binding}\sim \frac{m_Q}{2}\left(\frac{Y_Q^2}{4\pi}\right)^2\;,
\end{equation}
where $m_Q$ is quark mass, $Y_Q$ is its Yukawa coupling to the light
scalar (like in the SM, the SM4 quark masses are proportional to the
Yukawa constants, $m_Q\propto Y_Q$). The bound state of heavy quarks
will be much lighter than the sum of quark masses because of large
negative value of this binding energy, that can be even of the order
of the sum of the quark masses: \emph {e.g.} a meson-like \emph{light}
bound state of 6 $t$-quarks and 6 $t$-antiquarks was considered in
Ref.~\cite{Nielsen}.

A more economic configuration of heavy quarks bound by strong scalar
exchange is a bubble of the false vacuum $\phi=0$, surrounded by
domain wall, inside which massless fourth generation quarks are
confined. Indeed, in this case instead of quark masses the energy of
the false vacuum bubble contributes to the configuration mass. The
later may be much smaller than the sum of the quark masses if the
radius of the bubble is small.\footnote{The general idea of bound
  state lighter than constituents is not new: even massless bound
  states have been discussed in literature~\cite{Dimopoulos:1981xc}
  but within another setup.}  Below we describe a toy model where such
a configuration might be realized.

\section{Toy model}

We assume factorization of quark fields and the domain wall between
true and false vacua by localizing quarks inside a sphere of radius
$R$, where the coupled to quarks scalar $\phi$ (Higgs field) is fixed
to zero. We thus estimate the energy of the whole system as a sum of
two noninteracting parts: quarks and domain wall. Because of this
factorization, such toy model overestimates the total energy but is
sufficient for our purposes to illustrate the main idea we put
forward.

\begin{figure}[!htb]
\hspace*{-0.025\textwidth}
\begin{tabular}{ll}
\includegraphics[width=0.48\textwidth] {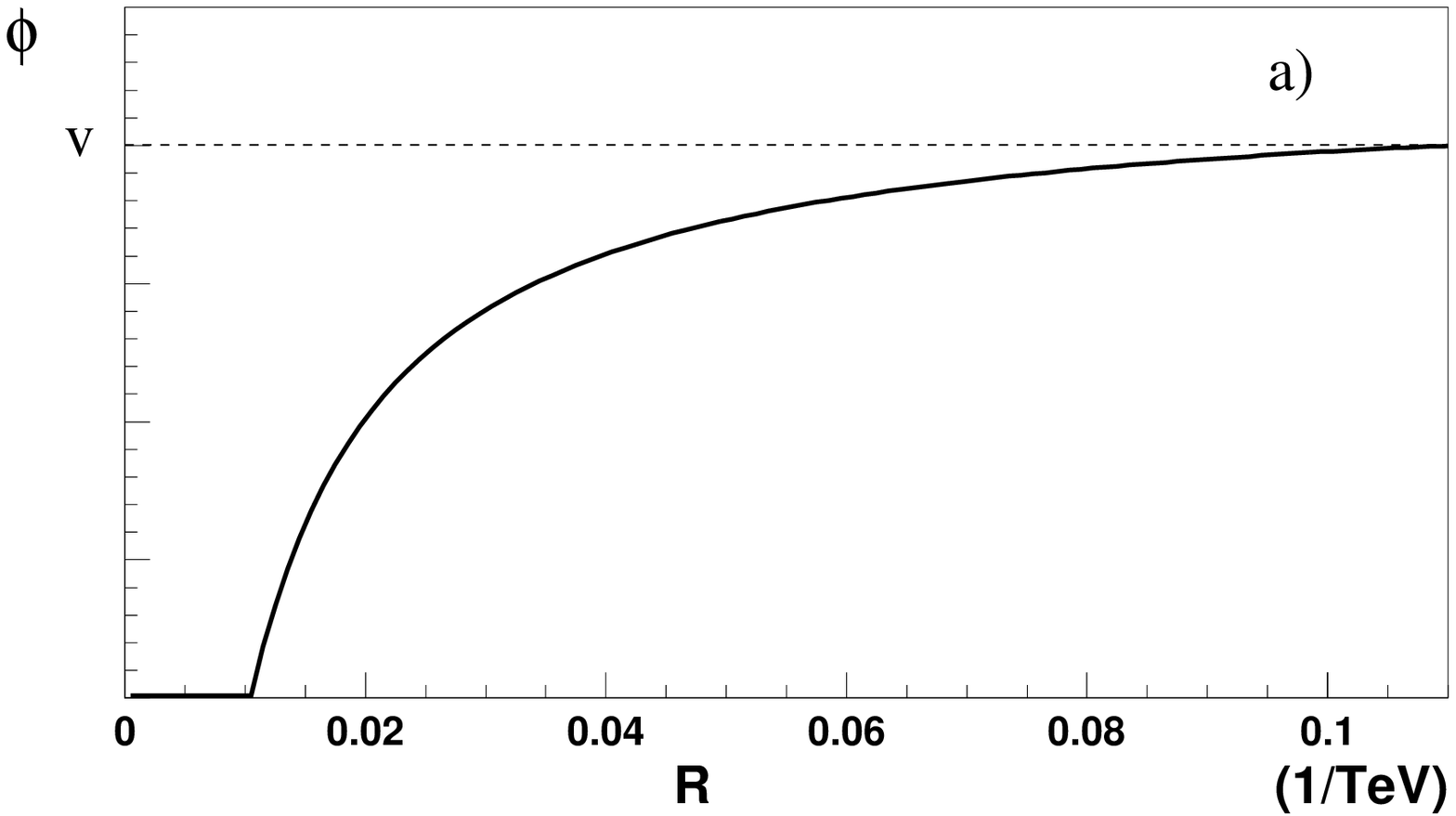}
&
\includegraphics[width=0.48\textwidth] {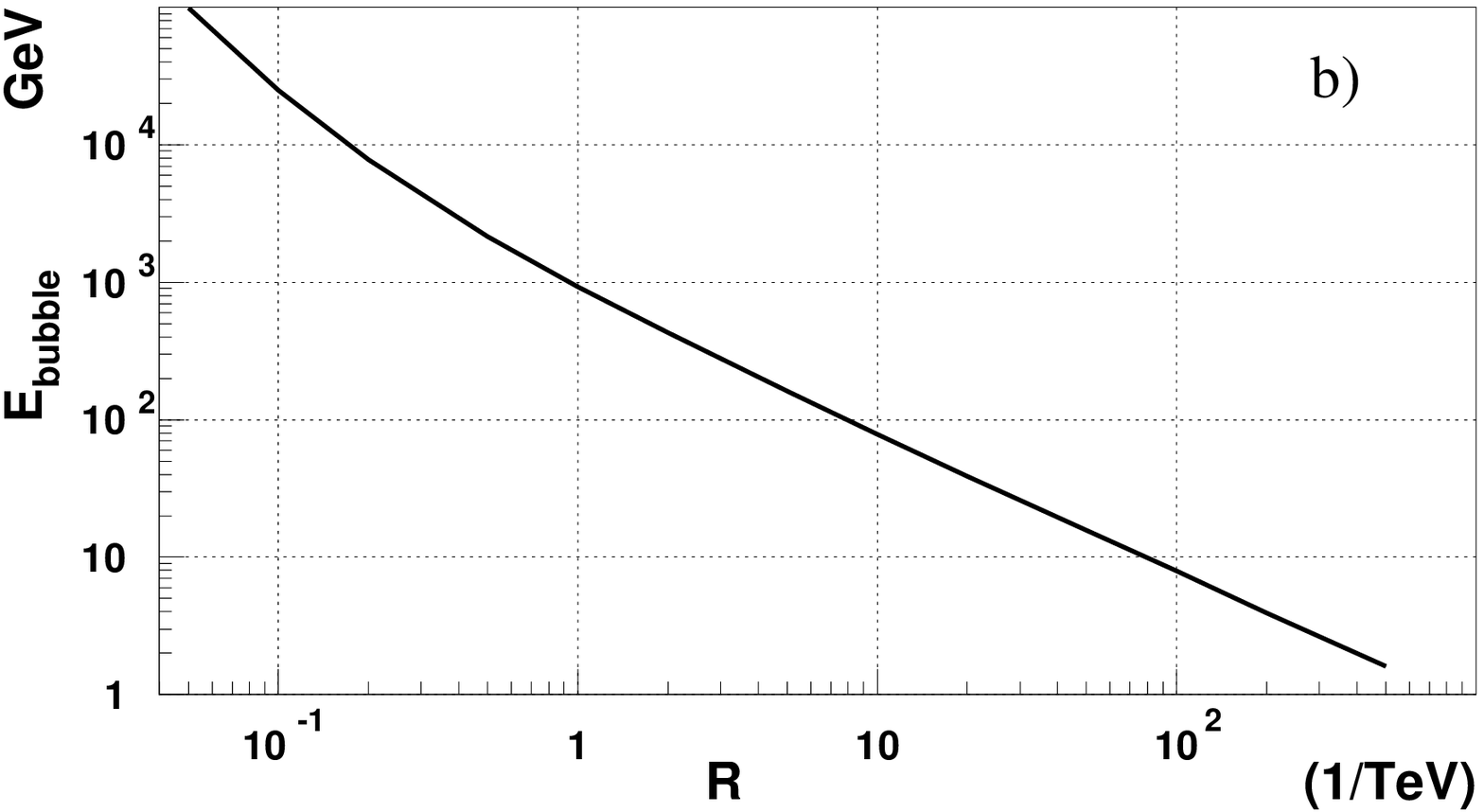}
\end{tabular}
\caption{ a) The scalar field $R$-dependence for inner radius fixed at
  $1/10\,$TeV$^{-1}$; b) Bubble energy dependence on inner radius.
\label{fields}}
\end{figure}

The simple spherically symmetric bubble profile for the scalar field
static configuration ($\phi=f(r)$) is schematically shown in
Fig.~\ref{fields}\,a) for $R=0.1\,$TeV$^{-1}$ as an example.
It is obtained numerically by minimizing the classical energy of the
scalar bubble, that includes the potential energy of false vacuum
inside the sphere, and both the potential and gradient energy in the
region of domain wall between false and true vacua:
\begin{equation}
\label{bubble-energy0}
E^{\rm bubble} = \frac{4}{3} \pi R^3 v^2 + \int_R^\infty \Big(
\frac{\lambda}{2}(f^2-v^2)^2 +
\big(\frac{\mathrm{d}f}{\mathrm{d}r}\big)^2 \Big) 4 \pi r^2 dr \; . \\
\end{equation}
For bubble radius $R$ smaller than inversed vacuum expectation value
of the SM Higgs field $v=246$\,GeV, $R<1/v$, the main contribution to
the energy comes from the domain wall rather than internal volume
(false vacuum energy). The optimal ``thickness'' of the domain wall is
roughly proportional to $R$, thus the total energy for small radius
turns out to be also proportional to the bubble radius (see
Fig.~\ref{fields}, b) as
\begin{equation}
\label{bubble-energy}
E^{\rm bubble}\approx 4.2 \pi\, v^2\, R\;.
\end{equation}

Inside the bubble quarks are massless, therefore only kinetic energy
$E^{\rm quark}_{\rm kinetic}\sim 1/R$ and potential energies due to
Yukawa and strong interactions contribute to total mass of quark
configuration in the bubble,
\begin{eqnarray}
E^{\rm quarks} \sim \left( 3 - 2\,\frac{Y^2}{4\pi}-3\,\alpha_s \right)
\,\frac{1}{R}\;.  \label{quark_en}
\end{eqnarray}
Here the first term in parenthesis is the kinetic energy of three
quarks inside the bubble, the second term comes from two pairs of
Yukawa interactions (between left and right quarks only, see below)
and the third term accounts for strong coupling in all three
pairs. For the moment, we ignore the contribution of $U(1)_Y$ and
$SU(2)_W$ gauge interactions, that can be both negative and positive
depending on the particular quark composition.

If the quark contribution~\ref{quark_en} to the total energy were
negative, which requires large values of Yukawa couplings $Y\gtrsim 4$
(not far from region where perturbative unitarity is lost,
$Y^2/(4\pi^2) \gtrsim 1$), the bubble is squeezed to a tiny size by
both domain wall pressure and scalar exchange interaction. The
infinite grip of the bubble may be prevented by the running
(decreasing) of Yukawa couplings with energy scale: Yukawas are
smaller at smaller radius. Then the bubble is stabilized at some $R$,
when
\begin{equation}
2\,\frac{Y^2(R)}{4\pi}+3\,\alpha_s(R) \approx 3\;.
\label{stabil}
\end{equation}
In this case the quarks contribution to the total energy is canceled
by equality of the potential and kinetic terms, and the baryon mass is
kept to be $\sim E^{\rm bubble}(R)$.

If reliable SM4 renormalization group calculations would exist in case
of large Yukawas, one could calculate the stabilization radius $R_*$
by solving
\begin{eqnarray}
\label{extremum}
\frac{d}{dR} \left( E^{\rm quarks}+E^{\rm bubble} \right)
\left|_{R=R_*}  \right.
&=0 \;, \nonumber \\
\frac{d^2}{dR^2} \left( E^{\rm quarks}+E^{\rm bubble} \right)
\left|_{R=R_*}  \right.
&>0\;.
\label{stability}
\end{eqnarray}
Introducing $\beta$-functions of the gauge and Yukawa
couplings as follows
\[
\beta_{\alpha_s}=\frac{d\alpha_s}{d\log\mu}\;,~~~ 
\beta_Y=\frac{d\left(Y^2/4\pi\right)}{d\log\mu}\;, 
\]
and setting energy scale at $\mu=1/R$ we obtain from
eq.\,\ref{extremum} for the value of $R_*$: 
\begin{equation}
\label{extremum-for-R}
4.2 \pi\, v^2\, R_*^2 = 3-2\,\frac{Y^2}{4\pi}-3\,\alpha_s-2\,\beta_Y
-3\,\beta_{\alpha_s}\;,
\end{equation}
where in the r.h.s. we omit the explicit dependence on $R_*$. For the
extremum at $R=R_*$ to be minimum one finds from~\ref{stability} 
\begin{equation}
\label{minimum-for-R}
4.2 \pi\, v^2\, R_*^2 >
2\,\beta_Y+3\,\beta_{\alpha_s}+2\,\beta_Y'+3\,\beta_{\alpha_s}' \;,
\end{equation}
where primes on r.h.s. refer to derivatives with respect to $\log\mu$
evaluated at $\mu=1/R_*$. If higher order contributions to
$\beta$-functions are subdominant, condition~\ref{minimum-for-R} is
fulfilled and the configuration is stable. Thus, at moderate values of
coupling constants the simple estimate we use would be enough.
However, for $Y \sim 4$-5 the SM4 renormalization group
equations\,\cite{renorm} suggest that both one-loop and two-loop
contributions are large and of the same order, but of the opposite
signs. This demonstrates that no reliable calculations can be done,
and even direction of running (falling or growing with energy scale)
is unknown. In addition the bare Yukawa constants are not known.
Thus, having an idea to explain DM by new antibaryons, we suppose that
\tp\ and \bp\ Yukawa couplings decrease with energy and
stabilization~\ref{stabil} is reached at $1/R\sim 100$\,TeV, hence the
antibaryon mass is near 5\,GeV, as needed. This requires the masses of
4-th generation to be $m_Q\gtrsim 700$\,GeV, close to the present
sensitivity of the direct searches at LHC~\cite{Chatrchyan:2012fp}.
We note, that perturbative calculations of the 4-th generation
contribution to electroweak observables~\cite{PDG} and the SM Higgs
boson production also become unreliable (see \emph{
  e.g.}~\cite{Chanowitz:2012nk}), and one might speculate they are
canceled. Otherwise, a new degree of freedom introduced to the model
may cure the situation and cancel the perturbative contributions of
the 4-th generation.

Although we can not predict the masses of the baryon family members
made of the 4-th generation quarks, some important properties can be
derived. In particular, the main feature is that the lightest baryons
turn out to be neutral and to have vanishing coupling to
$Z$-boson. Otherwise it would contribute to $Z$-boson width, see
Sec.\,\ref{Sec:Baryon-family}.

\section{Baryon family of 4-th generation quarks}
\label{Sec:Baryon-family}

Inside the false vacuum bubble the broken $SU(2)_W\times U(1)_Y$
symmetry is restored and for the proposed baryon candidates the
chirality of quarks matters. Baryons made of only lefthanded or only
righthanded quarks have no scalar coupling. We thus consider only
combinations of two lefthanded quarks and one righthanded quark, or
two righthanded quarks and one lefthanded quark: \qlp\qrp\qrp\ and
\qrp\qlp\qlp\ (\qp=\tp\ or \bp).

We should take into account corrections to the effective coupling of
quarks arising from the weak and hypercharge interactions. They change
the stabilization radius and therefore mass of the state. We list all
configurations with corresponding energies due to $U(1)_Y$ and
$SU(2)_W$ gauge interactions and couplings to photon and $Z$-boson in
Table~\ref{list}.
\begin{table}[!htb]
\caption{Baryon configurations with energies due to Yukawa ($\ayt
  \equiv \Yt^2/(4\pi)$, $\ayb\equiv \Yb^2/(4\pi)$), $U(1)_Y$ and
  $SU(2)_W$ gauge interactions and their effective couplings to
  photons ($Q^\gamma$, in units of proton electric charge), and
  $Z$-boson via vector and axial currents ($g_V^Z$ and $g_A^Z$, in
  units of $g_2$). Negative binding energy refers to repulsive
  forces. The DM candidate is anti-\brp\tlp\blp.}
\label{list}
\begin{center}
\begin{tabular}{l||r|r|r||r|r|r}
\hline 
baryon & Yukawa & $U(1)_Y$ & $SU(2)_W$ & $Q^{\gamma}$ & $g_V^Z$
& $g_A^Z$ \\ \hline

\blp\brp\brp & 2\ayb & $0$ & $0$ & $-1$ & $-\half+2s^2$ & $-\half$ \\

\blp\trp\brp & \ayt+\ayb & $\twt\aon$ & $0$ & 0 & $-\half$ & $-\half$
\\

\blp\trp\trp & 2\ayb & $-\eit\aon$ & $0$ & $+1$ & $-\half+2s^2$ &
$-\half$ \\

\tlp\brp\brp & 2\ayb & $0$ & $0$ & 0 & $\half$ & $\half$\\

\tlp\trp\brp & \ayt+\ayb & $\twt\aon$ & $0$ & $+1$ & $\half-2s^2$ &
$\half$\\

\tlp\trp\trp & 2\ayt & $-\fot\aon$ & $0$ & $+2$ & $\half-4s^2$ &
$\half$ \\

\brp\blp\blp & 2\ayb & $\ont\aon$ & $-\at$ & $-1$ & $-1+2s^2$ & $-1$\\

$\boldsymbol{b'}_{\!\!\!\boldsymbol{R}}
\boldsymbol{t'}_{\!\!\!\boldsymbol{L}}
\boldsymbol{b'}_{\!\!\!\boldsymbol{L}}$
& {\bf 2}$\boldsymbol{\ayb}$ 
& $\boldsymbol{\ont}\boldsymbol{\aon}$ & $\boldsymbol{\at}$ 
& $\mathbf{0}$ & $\mathbf{0}$ & $\mathbf{0}$  \\

\brp\tlp\tlp & 2\ayb & $\ont\aon$ & $-\at$ & $+1$ & $1-2s^2$ & $1$ \\

\trp\blp\blp & 2\ayt & $-\aon$ & $-\at$ & 0 & $-1$ & $-1$ \\

\trp\tlp\blp & 2\ayt & $-\aon$ & $\at$ & $+1$ & $-2s^2$ & 0 \\

\trp\tlp\tlp & 2\ayt & $-\aon$ & $-\at$ & $+2$ & $1-4s^2$ & $1$\\

\hline
\end{tabular}
\end{center}
\end{table}
From the LEP data the $Z$-boson partial width to invisible final state
except for three neutrino is below $6\times 10^{-4}$~\cite{PDG}, hence
``$Z$-charge'' of the lightest baryon (which is much lighter than half
of $Z$ mass, according to our arguments) should be exactly zero. There
is only one such a candidate in the Table\,\ref{list}:
\brp\tlp\blp\ baryon. It is also electrically neutral that is one of
the necessary conditions to be DM.  Other baryons (if exist) with
non-zero coupling to $Z$ should be heavier than $m_Z/2$.

One can expect that most combinations listed in the Table\,\ref{list}
do not form bound states. Eight out of the twelve listed candidates
include a pair of identical quarks. Such pairs should have symmetric
spin wavefunction, as two quarks are spatially symmetric, and color
antisymmetric; thus their total spin is 1, and there is a spin-spin
interaction energy between the remaining third quark and the
pair. Strength of the spin-spin interaction is not small in our case
of very compact state of relativistic components. Among the four
remaining configurations the \brp\tlp\blp\ and \trp\tlp\blp\ have
$SU(2)_W$ gauge force \emph{attractive}, hence reducing the bound
state masses. The former configuration (our DM) is naturally lighter,
as attractive $U(1)_Y$ force also reduces its mass, while repulsive
$U(1)_Y$ force increases the mass of charged baryon \trp\tlp\blp.

We conclude that natural candidate to be DM is \brp\tlp\blp. Its
binding energy is determined mostly by \bp\-quark Yukawa coupling \Yb,
while the mass of \trp\tlp\blp\; is determined by \tp-quark Yukawa
\Yt. Values of these Yukawas (and hence masses of \tp\ and \bp) have
to be chosen in such a way that cancellation of quark
energy~\ref{stabil} for \brp\tlp\blp\; happens at smaller radius $R$
(i.e. at higher energy scale for running couplings) than for
\trp\tlp\blp. Then \trp\tlp\blp\; is much heavier, since the energy of
the total configuration is then mostly the bubble
energy~\ref{bubble-energy}, which grows linearly with $R$. This
consideration suggests that \brp\tlp\blp\; is naturally the lightest
of all the baryon configurations listed in Table~\ref{list}, and
hereafter we suppose it is existing and its mass is about 5\,GeV.


\section{DM production and baryon asymmetry generation}

In the model the baryon asymmetry of the Universe as well as DM
component are produced after electroweak phase transition, when the SM
Higgs field gains nonzero vacuum expectation value and all quarks and
weak bosons become massive. Because of strong Yukawa couplings bound
states listed in Table\,\ref{list} start to form somewhat earlier.
When new phase has conquered the space everywhere except inside the
bound states, the latter become massive and decay into lighter bound
states and quarks. It is a nonequilibrium process and as a result the
required asymmetry between the lightest bound states \brp\tlp\blp\;
and $\bar{\brp} \bar{\tlp} \bar{\blp}$ may be generated with
approximately chosen CP-violating parameters in $4\times4$ extensions
of the CKM matrix. At the same process the rest of baryonic charge
transfers to the SM quarks. Then all \brp\tlp\blp\;-baryons annihilate
with antibaryonic counterparts leaving a requiring amount of dark
matter antibaryons made of $\bar{\brp} \bar{\tlp} \bar{\blp}$. Hence,
the model under discussion is capable of producing both DM and baryon
asymmetry of the Universe.


\section{Lifetime and interactions of the antibaryonic DM}

\begin{figure}[htb!]
\begin{center}
\hspace*{-0.025\textwidth}
\includegraphics[width=0.44\textwidth]{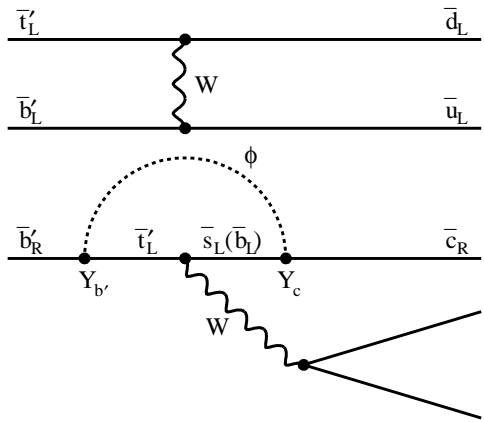}
\caption{Diagrams of DM antibaryon decay.}
\label{dec}
\end{center}
\end{figure}

If even the lightest ``Dark'' antibaryon has small mass of $\sim
5$\,GeV, it is not stable. Such antibaryons can decay to antiproton,
strange or charmed antibaryon via simultaneous transitions of all
heavy antiquarks to light antiquarks. Such non diagonal transition can
be provided only by weak interactions (a triple $W$-emission or
simultaneous $W$-exchange and $W$-emission), but one of the antiquark
is righthanded and more complicated diagram is required, like that
shown in Fig.~\ref{dec}.
It is extremely suppressed by CKM weak couplings to the first (second)
generation antiquarks and small Yukawa coupling of the second
generation. Thus, 5\,GeV antibaryon can (easily) be stable at
cosmological time scale.

DM particles scatter of each other with cross sections certainly
smaller than the geometric estimate $\sigma\sim 4\pi/R^2$. The number
is small enough to satisfy the upper limit from the Bullet cluster 1E
0657-56~\cite{Markevitch:2003at} and to ensure the stability of dark
matter at cosmological time scale in the interesting regions with
overdensity: inside galaxies and clusters. Likewise annihilation with
baryons is ineffective. However, the latter process is a signature of
the antibaryonic DM we propose: annihilation products in cosmic rays
may be searched for to test the model.

There are limits on elastic DM scattering off nuclei from direct
searches for Galactic DM at (underground) laboratories~\cite{PDG}.
Being decoupled from photon and $Z$-boson, the DM particles scatter
off ordinary matter with very small cross section. Also taking into
account small energy transfer due to small mass of 5\,GeV, one
concludes that the standard approach adopted in direct searches for
the dark matter particles, with energy release as the main signature,
is useless. More promising are experiments on searches for proton
decays: though in our model \emph{baryon number is perturbatively
  conserved,} the annihilation event of dark matter \emph{antibaryon}
with proton (neutron) will resemble baryon decay event very closely:
only energy release (total energy of the ``decay product'') is up to
5\,GeV.

\section{Conclusion}

In this Letter we have argued that DM may be represented by
antibaryonic matter, produced by very heavy quarks confined into
$M=5$\,GeV bound states, and baryogenesis can be provided without
baryon number violation. Neither new (conserved) charges nor
interactions are involved. A signature in a Super-K-like detector is
proton(neutron)-decay-like event but with more energetic outcoming
``decay''-products.  Analogous signature anticipated in cosmic rays is
annihilation with ordinary matter. A special study is worth to obtain
model predictions for cosmic ray experiments and direct searches.

4th generation of SM fermions must be searched for at LHC: expected
mass range is 700-900\,GeV, close to the present
limits~\cite{Chatrchyan:2012fp}. The dynamics we need to construct the
light antibaryons from heavy antiquarks may be provided by strong
coupling in Yukawa sector, though we have no reliable tools to
describe it quantitatively. Hopefully, this can be done using lattice
calculations. Note that configurations we operated with are very
compact: the antibaryon radius $R$ is much smaller than its naive
Compton wavelength $2\pi/M$, but is of order of the constituent
masses. One may hope to describe these configurations \emph{in
  perturbative regime} with new degrees of freedom introduced or by
extending the setup with extra spatial dimensions, while effective
Yukawa dynamics on the World 3-brane is at strong coupling.

\vskip 0.5cm 

{\bf Acknowledgments.} The authors are very thankful to A. Mironov,
A. Morozov and V. Rubakov for useful discussions. The work of D.G. is
supported in part by the grant of the President of the Russian
Federation NS-5590.2012.2, by MSE under contract \#8412 and by RFBR
grants 11-02-01528a, 13-02-01127a.

\end{document}